\documentclass[twocolumn,aps,pra,10pt,showpacs]{revtex4-1}
\usepackage{bm}
\usepackage{amsfonts}
\usepackage{amssymb}
\usepackage{amsmath}
\usepackage{graphicx}

\begin{document}

\title{Berry phase for spins of relativistic electrons}
\author{Iwo Bialynicki-Birula}\email{birula@cft.edu.pl}
\affiliation{Center for Theoretical Physics, Polish Academy of Sciences\\
Aleja Lotnik\'ow 32/46, 02-668 Warsaw, Poland}
\author{Zofia Bialynicka-Birula}
\affiliation{Institute of Physics, Polish Academy of Sciences\\
Aleja Lotnik\'ow 32/46, 02-668 Warsaw, Poland}

\begin{abstract}
Berry phase is a very general concept. It is applied here to families of solutions of the Dirac equation with different values of spin. The value of the Berry phase in the spin space is given by the same expression as was found before in the momentum space.
\end{abstract}

\maketitle

\section{Introduction}

In this paper we return to our studies of the Berry phase for relativistic particles. The very concept of the Berry phase is so general that it can be introduced, in principle, for every family of functions labeled by a set of continuously varying parameters. In our previous studies we identified the Berry phase for photons \cite{qed} and for electrons \cite{bb0}. In both cases the components of the momentum $\{p_x,p_y,p_z\}$ were chosen as the parameters. In the present study we find the Berry phase when the parameters $\{s_x,s_y,s_z\}$ are the components of the electron spin vector. Of course, certain properties of a family of wave functions become really interesting when one finds their application to experiments. This is why various occurrences of phases that were later linked with the Berry phase were not met with great interest. The essential ingredient introduced by Berry in \cite{berry} was the connection with the adiabatic changes of the wave function during the time evolution. The Berry phase is the {\em additional phase} that the wave function may acquire on top of the standard dynamical phase $\exp(-iEt)$.

\section{The Berry phase}

We start with a general definition of the Berry phase applicable to any quantum-mechanical system. Michael Berry in his first paper \cite{berry} wrote: ``Let the Hamiltonian ${\hat H}$ be changed by varying the parameters ${\bm R}=(X,Y,\dots)$ on which it depends. Then the excursion of the system between times $t=0$ and $t=T$ can be pictured as transport round the closed path ${\bm R}(t)$ in parameter space with Hamiltonian ${\hat H}({\bm R}(t))$ such that ${\hat H}({\bm R}(0))={\hat H}({\bm R}(T))$''. Even tough the Hamiltonian is essential for physical applications of these ideas, the very concept of Berry phase can be introduced without any reference to time evolution.

We use the Berry phase here as a purely geometrical general concept. Let us suppose that we have a quantum system whose wave functions $\Psi(q^i)$ depend on some parameters $q^1,q^2,\dots,q^n$. Let us define the $n$-dimensional vector field ${\bm V}(q^i)$ according to the formula
\begin{align}\label{v}
V_k(q^i)=i\langle\Psi(q^i)|\frac{\partial}{\partial q^k}\Psi(q^i)\rangle.
\end{align}
For a family of normalized wave functions, $\langle\Psi(q^i)|\Psi(q^i)\rangle$=1, the vector field $V_k(q^i)$ is real. Berry defined his ``geometrical phase'' as the following line integral evaluated along a closed contour $C$ in the space of parameters:
\begin{align}\label{c}
\gamma(C)=\oint_C{\bm V}(q^i)\!\cdot\!\bm{dq}.
\end{align}
At this level of abstraction, the time evolution does not play a role.

Berry considered first a three-dimensional space of parameters `` to employ familiar vector calculus'' and this will also be our case for a spinning electron. We can then apply the standard Stokes theorem and convert the line integral into an integral over the surface enclosed by the contour $C$,
\begin{align}\label{s}
\gamma(C)=\int\!\!\!\int_C\left({\bm\nabla}\times{\bm V}\right)\!\cdot\!\bm{dS},
\end{align}
where $\bm{dS}$ is the surface element.
We will apply this general approach here to relativistic electrons choosing the components of the spin $\{s_x,s_y,s_z\}$ as the parameters.

\section{Family of solutions of the\\Dirac equation}

Every solution of the Dirac equation in free space may be generated from an initial bispinor $\Psi({\bm r},0)$ by applying the time-evolution operator ($c=1,\hbar=1$),
\begin{align}\label{tev}
\Psi({\bm r},t)=e^{-iH_Dt}\Psi({\bm r},0),
\end{align}
where $H_D$ is the Dirac Hamiltonian. The action of the evolution operator can be effectively realized when the initial condition is represented as a Fourier integral,
\begin{align}\label{ft}
\Psi({\bm r},0)=\int\frac{d^3p}{(2\pi)^{3/2}E_p}e^{i{\bm p}\cdot{\bm r}}{\tilde{\Psi}}({\bm p}),
\end{align}
where $E_p=\sqrt{m^2+{\bm p}^2}$. In a relativistic theory it is convenient to use the invariant volume element $d^3p/E_p$ in momentum space. The Hamiltonian in momentum space becomes then a simple matrix, $H(\bm p)={\bm\alpha}\cdot{\bm p}+\beta m$, and we obtain:
\begin{align}\label{tev1}
\Psi({\bm r},t)=\int\frac{d^3p}{(2\pi)^{3/2}E_p}e^{-iH(\bm p)t}e^{i{\bm p}\cdot{\bm r}}{\tilde{\Psi}}({\bm p})
\end{align}
The $4\times 4$ matrix $H(\bm p)$ has four eigenvectors $u({\bm p},\pm)$ and $v({\bm p},\pm)$ which belong to doubly degenerate eigenvalues $E_p$ and $-E_p$, respectively. The eigenvectors can be defined by the following explicit formulas:
\begin{align}\label{uv}
u({\bm p},\pm)&=(\gamma^\mu p_\mu+m)a_\pm,\quad p_0=E_p\\
v({\bm p},\pm)&=(\gamma^\mu p_\mu+m)\tilde{a}_\pm,\quad p_0=-E_p,
\end{align}
where $a_\pm$ and $\tilde{a}_\pm$ are arbitrary, linearly independent bispinors. The Dirac matrices are chosen in the Weyl (chiral) representation,
\begin{align}\label{gamma}
\gamma^0=\left[\begin{array}{cc}0&I\\I&0\end{array}\right],\quad
\gamma^i=\left[\begin{array}{cc}0&-\sigma_i\\\sigma_i&0\end{array}\right].
\end{align}
The Fourier transform ${\tilde{\Psi}}({\bm p})$ expanded in the basis of the eigenvectors is:
\begin{align}\label{exp}
{\tilde{\Psi}}({\bm p})=\sum_{s=\pm}u({\bm p},s)f({\bm p},s)+\sum_{s=\pm}v({\bm p},s)g({\bm p},s).
\end{align}

The amplitudes $f({\bm p},s)$ describe the states of particles, while the amplitudes $g({\bm p},s)$ describe the states of antiparticles. Since we are interested here only in the states of electrons, we will assume that $g({\bm p},s)=0$. The action of $H(\bm p)$ on the  eigenvectors $u({\bm p},s)$ produces the eigenvalue $E_p$.  Therefore, the general solution of the Dirac equation describing electrons is the following superposition of plane waves:
\begin{align}\label{fourier}
\Psi({\bm r},t)=\sum_{s=\pm}\int\!\frac{d^3p}{(2\pi)^{3/2}E_p}u({\bm p},s)f({\bm p},s)e^{-ip\cdot x},
\end{align}
where $p\cdot x=E_pt-{\bm p}\!\cdot\!{\bm r}$.

Our aim is to describe the connection between the spin of relativistic electrons and the Berry phase. The detailed dependence of the amplitudes $f({\bm p},s)$ on momentum should not play a significant role. To disentangle the spin completely from the translational degrees of freedom we will assume that these amplitudes are equal to some spherically symmetric function $f(p)$ multiplied by complex coefficients $w_\pm$. We will choose the numerical bispinors $a_\pm$ in the form:
\begin{align}\label{w}
a_+=\{w_+,0,0,0\},\quad a_-=\{0,w_-,0,0\}.
\end{align}
Thus, the family of solutions of the Dirac equation which will be analyzed here is:
\begin{align}\label{fourier1}
&\Psi({\bm r},t)=\int\!\frac{d^3p}{(2\pi)^{3/2}E_p}f(p)e^{-ip\cdot x}\nonumber\\
&\times\left[
\begin{array}{c}m w_+\\m w_-\\(\sqrt{m^2+{\bm p}^2}-p_z)w_+-(p_x-ip_-)w_-\\-(p_x+ip_y)w_+(\sqrt{m^2+{\bm p}^2}+p_z)w_-\end{array}\right].
\end{align}
In the next Section we will determine the spin characteristics of these solutions.

\section{The spin vector}

The information about the spin of the electron is carried by the coefficients $w_\pm$. In order to identify this information we will evaluate the expectation value in the state (\ref{fourier1}) of the relativistic spin operator $\sigma_{\mu\nu}$,
\begin{align}
\sigma_{\mu\nu}&=
\frac{i}{2}\left(\gamma_\mu\gamma_\nu-\gamma_\nu\gamma_\mu\right),\\
\langle\sigma_{\mu\nu}\rangle&=\frac{\int\!dr^3\,{\bar\Psi}({\bm r},t)\sigma_{\mu\nu}\Psi({\bm r},t)}{\int\!dr^3\,{\bar\Psi}({\bm r},t)\Psi({\bm r},t)}.\label{spin}
\end{align}
The spatial components of this tensor $\hat{\bm s}=\{\sigma_{23},\sigma_{31},\sigma_{12}\}$ describe the ordinary spin. The substitution of (\ref{fourier}) into the definition (\ref{spin}), after the use of the Fourier representation, gives a remarkably simple result for $\hat{\bm s}$,
\begin{align}\label{spin1}
\langle\hat{\bm s}\rangle=\{s_x,s_y,s_z\}=\frac{{\bm w}^\dagger{\bm\sigma}{\bm w}}{{\bm w}^\dagger{\bm w}},
\end{align}
where  $\{\sigma_x,\sigma_y,\sigma_z\}$ are ordinary Pauli matrices and the Pauli spinor $\bm w$ is built from our two complex coefficients, $\bm w=\{w_+,w_-\}$. There is no trace of the function $f(p)$ in this formula. In this simple case the spin is completely decoupled from the translational degrees of freedom.

In order to use directly the prescription for the Berry phase in its simplest form, we invert the relation (\ref{spin1}) and we express $\bm w$ as a function of $\{s_x,s_y,s_z\}$. Since $\{s_x,s_y,s_z\}$ is a unit vector it can be represent in spherical coordinates in the form:
\begin{align}\label{spin2}
s_x=\cos\phi\sin\theta,\;s_y=\sin\phi\sin\theta,\;
s_z=\cos\theta.
\end{align}
In turn, a normalized spinor $\bm w$ which reproduces this vector in (\ref{spin1}) is:
\begin{align}\label{spin3}
{\bm w}=\{\cos(\theta/2)e^{-i\phi/2},\sin(\theta/2)e^{i\phi/2}\}.
\end{align}
In Cartesian coordinates, this formula reads:
\begin{align}\label{spin4}
{\bm w}=\left[\!\begin{array}{c}\sqrt{1\!+\!s_z/s}
\left(\sqrt{1\!+\!s_x/s_\perp}
\!-\!i\sqrt{1\!-\!s_x/s_\perp}\right)\\
\sqrt{1-s_x/s}\left(\sqrt{1\!+\!s_z/s_\perp}
+i\sqrt{1\!-\!s_x/s_\perp}\right)\end{array}\!\right],
\end{align}
where $s=\sqrt{s_x^2+s_y^2 +s_z^2}$ and $s_\perp=\sqrt{s_x^2+s_y^2}$. This formula for ${\bm w}$ holds only for $s_y>0$. For $s_y<0$ the spinor ${\bm w}$ must be replaced by its complex conjugate. After these preparations we are well prepared to tackle the problem of the Berry phase for relativistic electrons.

\section{The Berry phase and the spin\\of relativistic electrons}

The family of solutions of the Dirac equations parametrized by the components of the spin vector is obtained by substituting the expressions (\ref{spin4}) into the formula (\ref{fourier1}). We may now calculate the Berry phase following the procedures outlined in Section II. The evaluation of the scalar product (\ref{v}) is again most easily done with the use of the Fourier representation.

In order to generate a set of Dirac wave functions that depend on the the set of three parameters we substitute the spinor $\bm w$ expressed in terms of the spin vector $\bm s$ in (\ref{spin4}) into the formula (\ref{fourier1}). Next, we calculate the vector field ${\bm V}$ according to the definition (\ref{v}) where $\Psi(q^i)$ now is our solution of the Dirac equation. After tedious but straightforward calculations we obtain:
\begin{align}\label{f1}
{\bm V}=\frac{\{-s_ys_z ,s_xs_z,0\}}{(s_x^2+s_y^2)\sqrt{s_x^2+s_y^2+s_z^2}},
\end{align}
and the final result is:
\begin{align}\label{f2}
{\bm\nabla}\times{\bm V}=-\frac{\{s_x,s_y,s_z\}}{(s_x^2+s_y^2+s_z^2)^{3/2}}.
\end{align}
This expression has the same form as the formulas that were obtained in \cite{qed,bb0} for photons and electrons except that the momentum is now replaced by the spin. The value of the Berry phase (\ref{s}) is equal to the solid angle spanned by the contour $C$. To observe this manifestation of the Berry phase one has to force the electron spin to follow the path $C$. Such an experiment is analogous to the one with photons where the photons were forced to follow the path along the helical fiber \cite{tc}. In the case of electrons the role of the helical fiber is played by the magnetic field acting on the magnetic moment of the electron.

Our results may sound disappointing because the expression (\ref{spin1}) for the Dirac electron may be obtained in the nonrelativistic quantum mechanics based on the Pauli-Schr\"odinger equation. However, it is perhaps worth noticing, that in the simple case, when the spin is decoupled from the center of mass motion, the relativistic theory and the nonrelativistic one fully agree.

\end{document}